\documentclass[letterpaper,fleqn,usenatbib]{mnras}

\usepackage{enumerate}

\def\eg{{\it e.g.}}
\def\etal{{\it et al.}}
\def\etc{{\it etc.}}
\def\ie{{\it i.e.}}

\def\Msun{M$_\odot$}

\def\pmb#1{\setbox0=\hbox{$#1$}%
  \kern-0.25em\copy0\kern-\wd0
  \kern.05em\copy0\kern-\wd0
  \kern-0.025em\raise.0433em\box0}
\def\spmb#1{\setbox1=\hbox{${\scriptstyle #1}$}%
  \kern-0.25em\copy1\kern-\wd1
  \kern.05em\copy1\kern-\wd1
  \kern-0.025em\raise.0433em\box1}
\def\ba{\;\pmb{\mit a}}
\def\bv{\;\pmb{\mit v}}
\def\bx{\;\pmb{\mit x}}

\def\bI{\,\pmb{\mit I}}
\def\bJ{\,\pmb{\mit J}}
\def\bL{\,\pmb{\mit L}}
\long\def\Ignore#1{\relax}

\usepackage{graphicx}	% Including figure files
\usepackage{color}
\definecolor{red}{rgb}{0.7,0.1,0.1}
\definecolor{blue}{rgb}{0.2,0.2,0.8}
\definecolor{green}{rgb}{0.1,0.6,0.1}

\title[Particle selection]{Particle selection from an equilibrium DF}

\author[Sellwood]{J. A. Sellwood\thanks{E-mail:sellwood@as.arizona.edu} \\
Steward Observatory, University of Arizona, 933 N Cherry Ave,
  Tucson AZ 85722, USA}

\begin{document}
\label{firstpage}
\pagerange{\pageref{firstpage}--\pageref{lastpage}}
\maketitle

\begin{abstract}
When starting an $N$-body simulation of an isolated galaxy, it is
desirable to select particles from a distribution function to ensure
that the model is in equilibrium.  Random sampling from a DF is widely
used, but results in a set of particles that differs by shot noise
from that intended.  This paper presents a method to reduce sampling
noise that has been developed by the author in a many collaborations
over a number of years.  The technique has been partly described in
past papers, though the ideas have not previously been gathered
together, nor have its advantages been clearly demonstrated in past
work.  Of course, sampling errors can also be reduced by a brute force
increase in the number of particles, but methods to achieve the same
effect with fewer particles have obvious advantages.  Here we not only
describe the method, but also present three sets of simulations to
illustrate the practical advantages of reducing sampling error.  The
improvements are not dramatic, but are clearly worth having.
\end{abstract}


\section{Introduction}
\label{sec.intro}
Our understanding of the dynamics of isolated model galaxies has been
considerably advanced by $N$-body simulation, particularly of models
that begin from a settled state.  Yet creating an equilibrium set of
particles from which to start remains one of the most challenging
steps.  The simplest models are single component disks or spheroids in
which all the mass is in the particles, but one may wish to embed the
self-gravitating particles in an externally imposed, rigid
gravitational field or, in more elaborate models, to represent the
disk, bulge and halo as separate components in a combined equilibrium
model. Note that the gravitational potential in any model is the total
arising from all mass components, whether rigid or composed of mobile
particles, but the distribution function (DF) for each component must
be an equilibrium function in the total potential.

\citet{SA86}, \citet{Hern93}, \citet{KD95}, \citet{DS00}, \citet{HWK05},
\citet{WPB}, \citet{RAS09}, \citet{YS14}, and others have offered
techniques to create single or multi-component models, which achieve
something increasingly close to a global equilibrium.  Perhaps the
most sophisticated are the AGAMA models by \citet{Vasi19}, who uses
iterative techniques to devise equilibrium distribution functions
(DFs) for each component in the combined potential.  The DF is
generally expressed as a function of integrals $\{\bI\}$, such as the
classical energy and angular momentum $f(E,\bL)$ or actions $f(\bJ)$.

However, having found an equilibrium DF, many practitioners simply
select $N$ particles at random from the DF.  The procedure is to
generate candidate particles that are uniformly distributed in each
dimension of $(\bx,\bv)$-space, and then select only those for which
$f(\bI) > t f_{\rm max}$, with $t$ being a random value from a uniform
distribution $0 \leq t \leq 1$, and then keep trying until $N$ are
accepted.  Here, $f_{\rm max}$ is the largest value of $f(\bI)$, which
is usually for a particle at rest in the center of the
component.\footnote{The referee pointed out that $f_{\rm max}$ can be
infinite in rare cases, although it must be an integrable singularity,
since the mass within any small volume should be finite.  However,
random selection can still be achieved by transformation of variables,
such as we describe in \S\ref{sec.smooth}.}  There are many obvious,
and some clever, means to improve efficiency, such as limiting $|\bx|
\leq r_{\rm max}$ and $\bv$ so that the candidate particle at the
selected $\bx$ is gravitationally bound, \etc, but the vast majority
of candidate particles are rejected because $f(\bI)$ is generally much
smaller than its peak value over most of available phase space.  Not
only is this random sampling method inefficient, but it results in a
distribution of $\bI$ values that differs by shot noise from the
desired $f(\bI)$.  While shot noise declines as $N^{-1/2}$, the
benefit from increasing $N$ is painfully slow.

Random sampling works because $f$ specifies the mass in a
$2n$-dimensional volume element of Cartesian phase space $d^n\bx
d^n\bv$ and generating candidate particles that are uniformly
distributed in $(\bx,\bv)$ space leads naturally to the probability of
acceptance.  We could choose candidate particles in some other system
of coordinates, such as a set of integrals $\{\bI\}$, which requires
knowledge of the mass fraction, $d^m{\cal M}/d\bI^m$ in an
$m$-dimensional volume element of those integrals.  This function is
related to the mass in a Cartesian volume element through the Jacobian
determinant of the coordinate transform, and some examples for
different models are presented below.  (Note that the density in
action-angle coordinates is the same as in Cartesians, because the
transformation between the two systems is canonical, but those
variables suffer from the disadvantage that we generally do not have
simple or exact algebraic expressions for them.)  An important
advantage, though not the only one, of this approach is that the
dimensionality of the space of the variables $\bI$ is typically half,
or less, that of phase space, because we can neglect, at least while
we select the integrals, the corresponding phase angles, which must be
uniformly populated in an equilibrium model.

However, if we were to select values for the integrals, $\bI$, in the
same random manner as the coordinates, $(\bx,\bv)$, the distribution
of selected particles would still differ from the target DF by shot
noise and we would have gained little.  But the lower dimensionality
of integral space makes it possible to select values deterministically
in a smooth manner.  With knowledge of the mass in an element of
integral space, $d^m{\cal M}/d\bI^m$, we can divide the space of the
integrals into small boxes such that the size of each box
$\Delta^m\bI$ contains the mass fraction ${\cal M}/N$.  Then choosing
$\{\bI\}$ values for a single particle in each such box ensures that
selected particles have a density in the space of the integrals that
is as close as possible, for the finite number of particles, to
$f(\bI)$.

Rather than continue to discuss the general case, it is probably
easier to convey the principle of the technique in a few simple
examples, and we begin with the case of a razor-thin disk.  Some extra
effort is required to write the code to select a smooth distribution
of integrals, but the running time to generate a given number of
particles is comparable, or sometimes shorter, than even an optimized
random sampling method.

There are two distinct strategies to limit Poisson noise.
\S\ref{sec.smooth} describes the first, which is to reduce sampling
errors in the selection of particles from the DF.  The second is the
much simpler strategy of imposing near axial symmetry, which we
outline in \S\ref{sec.qstart}.  While axial symmetry results in a
dramatic improvement, the more modest benefits of investing the extra
effort in the selection of particles is still worthwhile.  Note, they
are two independent techniques that can be employed separately, as we
illustrate in \S\ref{sec.results}.

\section{Reduction of sampling error}
\label{sec.smooth}
The selection of particles in multi-component models should be made
one component at a time.  We here provide a detailed description of
the technique, which may seem labored, in order that every step is
clear.

\subsection{A 2D disk with $f(E,L_z)$}
\label{sec.2Ddisk}
Following \citet{Kaln76}, we write the disk surface density as
\begin{equation}
\Sigma(R) = 2 \int_{v_R=0}^{\rm max} \int_{v_\phi=0}^{\rm max} f(v_R,v_\phi)\big|_R \; dv_\phi dv_R,
\label{eq.surfd}
\end{equation}
where the factor 2 arises from having omitted inwardly moving stars
from the outer integral.  Also retrograde stars are generally omitted
from the DF in rotationally-supported disks, but the equilibrium is
unaffected if the sign of $v_\phi$ is later reversed for some, which
is desirable in order to smooth an unphysical discontinuity in the DF
at $v_\phi=0$.

We assume an axisymmetric potential $\Phi(R)$, so that $v_\phi =
L_z/R$ and $E=\Phi(R) + {1 \over 2} (v_R^2 + L_z^2/R^2)$, which we
rearrange to obtain $v_R = [2(E-\Phi) - (L_z/R)^2]^{1/2}$, and
change variables from $(v_R,v_\phi)$ to $(E,L_z)$
\begin{equation}
\Sigma(R) = 2\int\int {\partial(v_R,v_\phi) \over \partial(E,L_z)} f(E,L_z) \;dL_z\,dE.
\end{equation}
The determinant of the Jacobian matrix is
\begin{equation}
{\partial(v_R,v_\phi) \over \partial(E,L_z)} = \left| \matrix{ 1/v_R & 0 \cr
  -L_z/(R^2v_R) & 1/R} \right| = {1 \over Rv_R}.
\end{equation}
\Ignore{since the partial derivatives are:
\begin{eqnarray}
& \displaystyle \left.{\partial v_R \over \partial E}\right|_{L_z} = {1 \over v_R}, \quad
& \left.{\partial v_\phi \over \partial E}\right|_{L_z} = 0, \nonumber \\
& \displaystyle \left.{\partial v_R \over \partial L_z}\right|_E = -{L_z \over R^2 v_R}, \quad
& \left.{\partial v_\phi \over \partial L_z}\right|_E ={1 \over R}.
\end{eqnarray}}
We therefore find
\begin{equation}
\Sigma(R) = 2\int_{\Phi(R)}^0 \int_0^{R\{2[E-\Phi(R)]\}^{1/2}} {f(E,L_z) \over Rv_R} \;dL_z\,dE,
\end{equation}
where the upper limit on the inner integral is $L_z$ of a circular
orbit at radius $R$.

\Ignore{
For his own purposes, \citet{Kaln76} wrote his eq.~(3) as
\begin{equation}
\Sigma(R) = 2\int\int {E f(E,L_z) \over [(1-E/\Phi)(-2R^2\Phi/L_z^2) - 1]^{1/2}} {dE \over E}{dL_z \over L_z}, \nonumber
\end{equation}
``with E varying between the limits $\Phi$ and 0, and the range of
$L_z$ restricted between 0 and $[(1 - E/\Phi)t]^{1/2}$, where
$t \equiv (-2R^2\Phi)^{1/2}$.''  We factor $t$ into the upper
integration limit to find $L_{z,\rm max}(E) = R[2(E-\Phi)]^{1/2}$ and
manipulate the equation by first multiplying out the factors in the
denominator obtaining
\begin{equation}
\Sigma(R) = 2\int_\Phi^0\int_0^{L_{z,\rm max}(E)} {E L_z f(E,L_z) \over R[2(E-\Phi) - L_z^2/R^2]^{1/2}} {dL_z \over L_z}{dE \over E}, \nonumber
\end{equation}
which, recognizing $v_R = [2(E-\Phi) - L_z^2/R^2]^{1/2}$, agrees with eq.~(5) above.
}

The total mass of the axisymmetric disk is
\begin{eqnarray}
& \displaystyle {\cal M} = 2\pi\int_0^\infty R\Sigma(R)\; dR \nonumber \\
& \displaystyle = 4\pi \int_0^\infty \int_{\Phi(R)}^0\int_0^{R\{2[E-\Phi(R)]\}^{1/2}} {f(E,L_z) \over v_R} \; dL_z\,dE\,dR.
\end{eqnarray}
Interchanging the order of integration, we obtain
\begin{equation}
{\cal M} = 4\pi \int_{\Phi(0)}^0 \int_0^{L_{z,\rm
max}(E)} \left\{\int_{R_{\rm peri}}^{R_{\rm apo}} {dR \over v_R}\right\} f(E,L_z) \;dL_z\,dE,
\label{eq.totM}
\end{equation}
where $R_{\rm peri}$ and $R_{\rm apo}$ are respectively the radii of
the inner and outer roots of $[2(E-\Phi) - L_z^2/R^2]^{1/2} = 0$,
where the radial velocity changes sign, and $L_{z,\rm max}(E)$ is
the angular momentum of a circular orbit of energy $E$.  Note
that $f$ is independent of $R$, so the inner integral in braces, which
is over outwardly moving stars only (eq.~\ref{eq.surfd}), is
\begin{equation}
\int_{R_{\rm peri}}^{R_{\rm apo}} {dR \over v_R} = {\tau(E,L_z) \over 2},
\end{equation}
where $\tau$ is the full radial oscillation period for a particle in the
adopted total potential.  On differentiating the expression for ${\cal
  M}$, we obtain the mass in an infinitesimal element of $(E,L_z)$ space:
\begin{equation}
{d^2{\cal M} \over dEdL_z} = 2\pi \tau(E,L_z)f(E,L_z).
\end{equation}

\subsubsection{Slicing by integrals}
We use this last equation to slice $(E,L_z)$ space into finite
elements.  The mass as a function of $E$ is
\begin{equation}
M(E) = \int_{\Phi(0)}^E m_E(L_{z, \rm max}) \;dE,
\label{eq.sliceE}
\end{equation}
with
\begin{equation}
m_E(L_z) = 2\pi \int_0^{L_z} \tau(E,L_z) f(E,L_z) \;dL_z.
\label{eq.sliceL}
\end{equation}
The functions $m_E(L_z)$ and $M(E)$ are known only numerically, in
general potentials, but they are monotonically increasing functions of
their arguments since both $f$ and $\tau$ are positive.  Thus we can
determine the values $E(i_E)$ that divide ${\cal M}$ into $j_E$ equal
pieces, $\delta M(E) = {\cal M}/j_E$, as can the values $L_{z,E}(i_L)$
that divide $\delta M(E)$ into $j_L$ equal pieces.  The mass enclosed
in the rectangle bounded by $E(i_E)$ and $E(i_E +1)$ and
$L_{z,E}(i_L)$ and $L_{z,E}(i_L +1)$ is ${\cal M}/(j_Ej_L)$.  If we
were to choose just one particle to have $(E,L_z)$ values that lie in
this rectangle, we would reduce the sampling error of the function
$f(E,L_z)$ to the minimum possible for the finite number, $j_Ej_L$, of
particles.

Since the product $j_Ej_L$ is large, the variation of $f(\bI)$ over
this small volume $\Delta^m\bI$ is generally small and we select
values for the integrals $(E,L_z)$ to be those at a randomly chosen
point within each mass element, which adds a little random noise to
our careful sampling technique.  The reasons for a random selection
are that it is probably not a good idea for the selected particles to
lie in a regular lattice in integral space and, perhaps more
importantly, that $f(E,L_z)$ rises steeply in disks as $L_z
\rightarrow L_{z, \rm max}(E)$, and choosing the center of every box,
say, would introduce a bias against closely circular orbits.

\begin{figure}
\includegraphics[width=\hsize,angle=0]{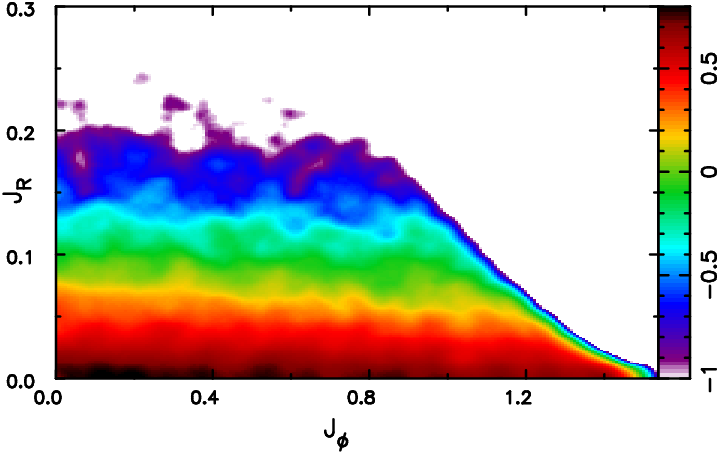}
% jasmine:/data/sellwood/5500/5504/dflook.s run5499
\includegraphics[width=\hsize,angle=0]{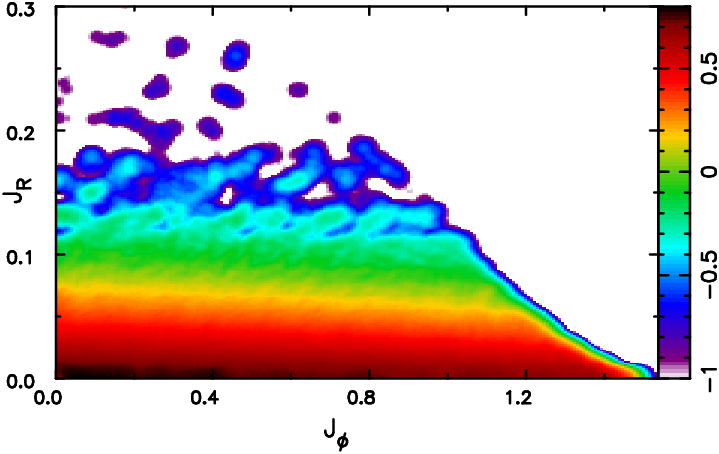}
% jasmine:/data/sellwood/5500/5504/dflook.s run5505
\caption{The logarithmic density of selected particles in the space of
  the two actions $J_\phi \equiv L_z$ and the radial action $J_R$ when
  random sampling is used (upper panel) and when $E$ and $L_z$ are
  selected smoothly, as described here.  $40$K particles were selected
  by each method from the isochrone/12 DF described in
  \S\ref{sec.modetest}, having no active particles outside $R=4$. The
  sloping boundary at high $L_z$ is caused by eliminating particles
  having enough energy to cross that radius.  Notice the
  non-smoothness in the densely populated lower part of the upper
  panel.}
\label{fig.DFact}
\end{figure}

The benefit of smooth selection is illustrated in
Fig.~\ref{fig.DFact}, where the density of selected particles in the
densely populated region at small $J_R$ is smoother in the lower panel
than when random sampling is used (upper panel).  These
non-uniformities resulting from random sampling contribute to the
evolution, and effectively create a different DF whose dynamical
properties diverge from that of the intended smooth case.

\subsubsection{Unequal mass particles}
\label{sec.uqmass}
The total number of particles in a component is $N =j_Ej_Lj_Rj_\phi$,
with $j_R$ and $j_\phi$ defined below.  If all particles have equal
mass, each has the mass ${\cal M}/N$.

However, it is sometimes useful to employ particles having a range of
masses in order to concentrate more low mass particles in some part of
integral space, which can readily be accomplished by this technique.
The relative particle masses can be varied as the weight function
$w(E,L_z)$, which must be positive over the entire range of these
integrals, and which should be divided into eq.~(\ref{eq.totM}) to
yield a pseudo-total mass
\begin{equation}
{\cal M}^\prime = 2\pi \int_{\Phi(0)}^0 \int_0^{L_{z,\rm max}(E)}
{\tau(E,L_z) f(E,L_z) \over w(E,L_z)} \;dL_z\,dE.
\label{eq.totpM}
\end{equation}
We also divide the integrand in eq.~(\ref{eq.sliceL}) by the function
$w(E,L_z)$ in order to make the appropriate changes to the values of
$E(i_E)$ and $L_{z,E}(i_L)$ that slice up ${\cal M}^\prime$.  Then the
mass of each of the $j_Rj_\phi$ particles that have these revised
integrals is $w(E,L_z){\cal M}^\prime/N$.  Note that because of this
factorization, there is no need to normalize the function $w(E,L_z)$.

The same idea can be applied to particle selection for non-disk
components discussed in this section.

\subsubsection{Selecting coordinates}
Having selected the integrals, $E$ and $L_z$, which define an orbit in
the potential of the adopted model, we then need to choose positions
and velocities.  Integrating the orbit of the selected $(E,L_z)$ for
half a radial period in the adopted potential determines the time
($\propto v_R^{-1}$) a particle would spend at each radius.  We select
$j_R$ values of $R$ at random from uniform fractions {\em in time} of
the orbit half-period.  (I have experimented with spacing these radii
at equal time intervals in order to further reduce jitter in the
radial mass profile, but with no detectable improvement in practice.)

For each $R$, we have $v_\phi = L_z/R$, and $v_R = \pm[2(E-\Phi) -
  L_z^2/R^2]^{1/2}$, with either sign having equal probability since
the inward and outward motions are anti-symmetric.  The last remaining
coordinate to be determined is the azimuthal position $\phi$, which
can be chosen at random from a uniform distribution, in which case
$j_\phi = 1$.  If $j_\phi>1$, we can place copies of the particle each
having coordinates $(R, v_R, v_\phi)$ at different azimuths $\phi$,
and substantially reduce the effects of shot noise by spacing them
regularly in $\phi$ for a quiet start \citep{Sell83, SA86}, which is
described in detail in \S\ref{sec.qstart}.

Random selection of azimuthal phases really does introduce shot noise
into the initial density distribution, that may be readily suppressed
by imposing initial axial symmetry.  But we also randomly select the
radial phase and introduce a small random element in the selection of
$(E,L_z)$ values, which indeed re-introduces some noise and may seen
to negate all the advantages gained in the effort to make a smooth
selection of integral values.  We address this point in
\S\ref{sec.modetest}.

\subsection{A thickened disk}
\label{sec.3Ddisk}
DFs that are functions of $E$ and $L_z$ only are unsuitable for
thickened disks because the velocity dispersions in the radial and
vertical directions are equal \citep[][eq.~13]{Sell14b}, making the
disk unrealistically thick.  However, we are generally unable to
construct DFs that are functions of three integrals, $f(E,L_z,I_3)$
for at least two compelling reasons.  First, we do not have a simple
expression for $I_3$, aside from numerical approximations
\citep{Binn10} and, second, parts of phase space can be chaotic.
Despite this, \citet{Vasi19} provides two approximate DFs for
thickened disks that are functions of 3 actions.

A more general and workable approximate method to construct an
equilibrium for a 3D disk is to employ a 2D equilibrium model for a thin
disk and treat the vertical motion as a separate 1D problem.  We
integrate the vertical 1D Jeans equation \citep[][eq.~422b]{BT08} for
a slab, \ie\ neglecting radial variations.  In this case
\begin{equation}
\sigma_z^2(R,z) = {1 \over \rho(R,z)} \int_z^\infty \rho(R,z^\prime)
      {\partial \Phi \over \partial z} \; dz^\prime,
\label{eq.vert1D}      
\end{equation}
where the vertical gradient of the total potential $\Phi$ should be
determined from the disk itself, as well as any additional mass
components.  This formula generally yields an acceptable equilibrium
when the radial excursions of disk particles are small, but when this
is not the case, the disk adjusts quickly to a mild imbalance because
the vertical oscillation period is short -- see \S\ref{sec.diskhalo}
for an example.  The method is versatile because it allows any
reasonable vertical density profile $\rho(R,z)$, and works in the
presence of other mass components that contribute to the total
potential.

\citet{SB16} reviewed this and other methods to construct a thickened
disk, and found that methods based on a St\"ackel approximation are
superior and yield a better disk equilibrium.  We report in
\S\ref{sec.diskhalo} that the initial model we use there is slightly
out of balance, and a better method would be desirable, especially for
disks that might be hotter or thicker than that we employed.

\subsection{A spherical model with $f(E,L)$}
\label{sec.ansph}
A sphere of stars that has an anisotropic velocity distribution,
requires a DF that is a function of both $E$ and the total angular
momentum $L$.  The velocity dispersion tensor must everywhere be
aligned with the radius vector, else the mass distribution will be
aspherical.  Since all orbits are planar, the velocity at each point
has just two components, $v_r$ and $v_\perp$, and the $v_\perp$
directions of all orbits passing through any point are uniformly
distributed in a circularly symmetric fashion about the radius vector.
Thus we have
\begin{equation}
\rho(r) = \int f(v_r,v_\perp)\big|_r d^3\bv = \int\int 2\pi v_\perp f(v_r,v_\perp)\big|_r \;dv_\perp\,dv_r,
\end{equation}
where the $2\pi v_\perp$ factor arises from integrating out the uniform
directional distribution of $v_\perp$.  As usual, $L=rv_\perp$ and $E
= \Phi(r) + {1 \over 2} (v_r^2 + v_\perp^2)$, so $v_r=[2(E-\Phi) -
(L/r)^2]^{1/2}$.

Changing variables from $(v_r,v_\perp)$ to $(E,L)$, we have
\begin{equation}
\rho(r) = 4\pi\int\int {\partial(v_r,v_\perp) \over \partial(E,L)}
v_\perp f(E,L) \;dL\,dE,
\end{equation}
where the integral is doubled, as for the disk, to take account of
both inwardly and outwardly moving stars. The determinant of the
Jacobian matrix clearly is $1/(rv_r)$, since it differs from that in
the disk simply by the substitution of $v_\perp$ for $v_\phi$, and we
therefore find
\begin{equation}
\rho(r) = 4\pi\int\int {v_\perp \over rv_r} f(E,L) \;dL\,dE.
\end{equation}
The total mass of the sphere is therefore
\begin{eqnarray}
{\cal M} & = & 16\pi^2\int_0^\infty r^2 \int_{\Phi(r)}^0\int_0^{\rm max}
{v_\perp \over r v_r} f(E,L)\;dL\,dE\,dr \nonumber \\
& = & 8\pi^2 \int_{\Phi(0)}^0\int_0^{L_{\rm max}(E)} L \tau(E,L) f(E,L)\;dL\,dE,
\end{eqnarray}
where $\tau$ is the full radial oscillation period, as before. Thus
the mass in an infinitesimal element of $(E,L)$ space is
\begin{equation}
{d^2{\cal M} \over dEdL} = 8\pi^2 L \tau(E,L)f(E,L).
\end{equation}

\subsubsection{Particle selection}
The procedure is very nearly the same as for the disk case described
in \S\ref{sec.2Ddisk} above.  We use this last equation to slice
$(E,L)$ space into elements that each contain mass ${\cal
M}/(j_Ej_L)$, and then select values for the integrals of each
particle $(E,L)$ to be those at a randomly chosen point within each
element.  We again integrate the planar orbit of the selected $(E,L)$
for half a radial period in the adopted spherical potential to
determine the time a particle would spend at each radius, and we
select $j_r$ values of $r$ (typically at random) from uniform fraction
of the radial half-period.  Having chosen $r$, we have $v_\perp = L/r$,
and $v_r = \pm[2(E-\Phi) - L^2/r^2]^{1/2}$, again with either sign
having equal probability.

In this case, we must also choose the orientation of the orbit plane,
which is uniformly distributed in the cosine of the polar angle
$-\pi/2 < \theta < \pi/2$, and which requires $r$ to be resolved into
$(R,z)$ and $v_r$ into components $v_R$ and $v_z$.  The last remaining
coordinate to be determined is the azimuthal position $\phi$ -- see
\S\ref{sec.qstart}.

\goodbreak
\subsection{An ergodic sphere model with $f(E)$}
In this case, the velocity distribution is everywhere isotropic and we
have $d\bv^3 = 4\pi v^2dv$.  Thus
\begin{equation}
\rho(r) = 4\pi \int_0^{v_{\rm max}} f(E) v^2 dv,
\end{equation}
where $v_{\rm max} = \{2[E-\Phi(r)]\}^{1/2}$.  While $f(E)$ is
independent of $L$, the radial period $\tau(E,L)$ is not, and it is
therefore simplest to adopt the same procedure to select integrals
$(E,L)$ and particle coordinates as for the anisotropic, spherical
case in \S\ref{sec.ansph} above, while keeping the DF uniform in $L$.

\subsection{A spheroidal model with $f(E,L_z)$}
\label{sec.sphrd}
For this case, we choose the $z$-axis as the axis of rotational
symmetry, so that $\rho$ and $\Phi$ are functions of both $R$ and $z$.
The velocity at any point is $(v_\phi,\psi,v_m)$
\citep[][\S4.4.1]{BT08}, with $v_\phi$ being perpendicular to the
radius vector and lying in a plane at height $z$ parallel to the
symmetry plane.  The component $v_m$, lies in the meridional
plane\footnote{A plane containing the star that rotates about the
symmetry axis, at the time-varying rate $\dot\phi = L_z/R$.  The
Cartesian position of the star within the plane at any instant is
$(R,z)$.} and is uniformly distributed in the angle $\psi$ such that
$v_R = v_m\cos\psi$ and $v_z = v_m\sin\psi$.  Therefore
\begin{eqnarray}
\rho(R,z) & = & \int f(\bx,\bv) \; v_m\,dv_m\,d\psi\,dv_\phi \nonumber \\
& = & 2\pi \int_0^{\sqrt{2(E-\Phi)}} \int_{-v_{\phi, \rm max}}^{v_{\phi, \rm max}} f \; v_m\,dv_\phi\,dv_m,
\end{eqnarray}
since $f$ does not depend on $\psi$, and $v_{\phi, \rm max} =
[2(E-\Phi)-v_m^2]^{1/2}$.  Note that $v_m \geq 0$ while
$v_\phi$ can have either sign.  Changing variables,  the determinant of
the Jacobian matrix, ${\partial(v_m,v_\phi) / \partial(E,L_z)} =
(Rv_m)^{-1}$, and so we have
\begin{equation}
\rho(R,z) = 2\pi \int_{\Phi(R,z)}^{E_{\rm max}} \int_{-L_{z,\rm max}(E)}^{L_{z,\rm max}(E)} {f(E,L_z) \over Rv_m}v_m\,dL_z\,dE,
\end{equation}
where $L_{z,\rm max}(E) = R[2(E-\Phi)]^{1/2}$.

\setbox1\vbox{
The mass of the spheroid is
\begin{eqnarray}
{\cal M} & = & 2\pi \int \int \rho(R,z)R \;dz\,dR \nonumber \\
& = & 4\pi^2 \int_0^\infty \int_{-z_{\rm max}(R)}^{z_{\rm max}(R)} \int_{\Phi(R,z)}^0 \int_{-L_{z,\rm max}(E)}^{L_{z,\rm max}(E)}  \nonumber \\
\nobreak
& & \qquad\qquad\qquad\qquad\phantom{\int_0^1} f(E,L_z) \;dL_z\,dE\,dz\,dR \nonumber \\
& = & 4\pi^2 \int_{\Phi(0,0)}^0 \int_{-L_{z,\rm max}(E)}^{L_{z,\rm max}(E)} \left\{ \int_{\rm peri}^{\rm apo} \int_{-z_{\rm max}(R)}^{z_{\rm max}(R)}\;dz\,dR\right\} \nonumber \\
& & \qquad\qquad\qquad\qquad\qquad\phantom{\int_0^1} f(E,L_z)\;dL_z\,dE.
\end{eqnarray}
}

\box1
\noindent  Note that the boundary of the double integral within the
braces is the zero velocity curve in the meridional plane of an orbit
having integrals $(E,L_z)$ and, since the integrand is unity within
that boundary, the double integral is simply the area $S(E,L_z)$
bounded by the zero-velocity curve. Thus the mass in an infinitesimal
element of $(E,L_z)$ space is
\begin{equation}
{d^2{\cal M} \over dEdL_z} = 4\pi^2 S(E,L_z)f(E,L_z).
\end{equation}

Once again, we use this last equation to slice $(E,L_z)$ space into
elements that each contain mass ${\cal M}/{j_Ej_L}$, and then select
values for the integrals $(E,L_z)$ to be those at a randomly chosen
point within each element.  Since the area $S$ in the meridional plane
is uniformly populated, we choose $j_R$ values of both $R$ and $z$ at
random from within that area.  The selected radii determine $v_\phi =
L_z/R$ and $v_m = \{2[E-\Phi(R,z)] - L_z^2/R^2\}^{1/2}$ (strictly
positive this time).  The velocity $v_m$ is oriented at random in the
meridional plane so we choose $0<\psi<2\pi$, from which $v_R$ and
$v_z$ follow.  As always, it remains only to choose the azimuth
$\phi$.

\begin{figure}
\includegraphics[width=\hsize,angle=0]{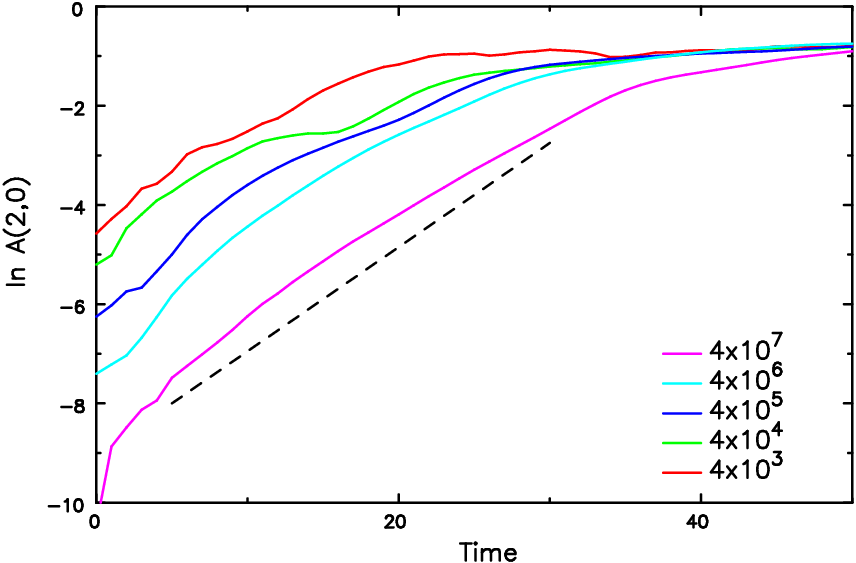}
% jasmine:/data/sellwood/5500/5503/amplot.s
\includegraphics[width=\hsize,angle=0]{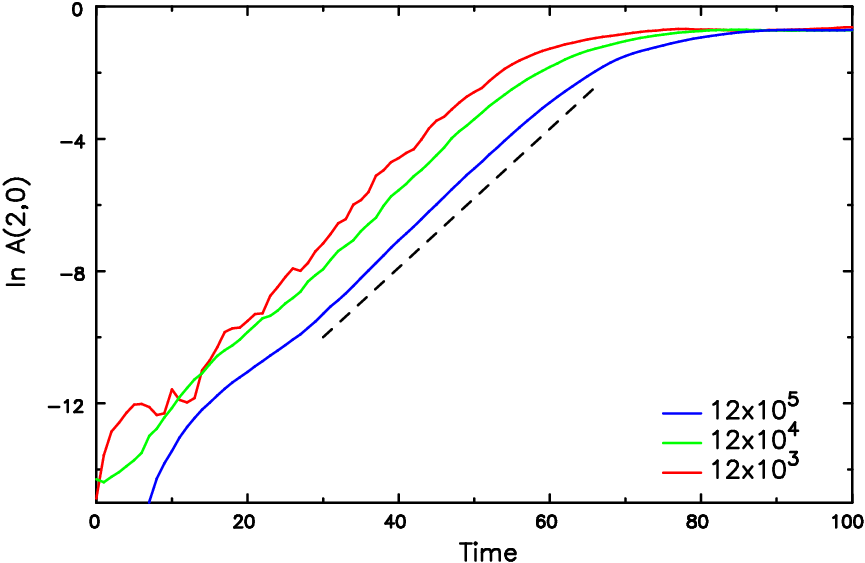}
% jasmine:/data/sellwood/5500/5505/amplotQ.s
\caption{Upper: The growth of a bisymmetric disturbance in an unstable
  disk in five simulations having differing numbers of particles.  The
  unstable disk model and the simulation method are described in
  \S\ref{sec.modetest} and all began with the particles placed at
  random azimuths.  The dashed line indicates the expected growth rate
  of the dominant instability. Lower: As for the upper panel, but for
  three simulations that began with two additional replica particles
  placed evenly around rings for each case. Note the changes of scale
  to both axes.}
\label{fig.amplot}
\end{figure}

\section{Azimuthal symmetry}%{Quiet starts}
\label{sec.qstart}
If the azimuthal coordinates $\phi$ were selected at random, then
non-axisymmetric forces would be subject to the full level of shot
noise expected from $N$ particles.  In an unstable model, the seed
amplitude of a non-axisymmetric instability would be high, unless $N$
were extremely large, and by the time linear growth causes the
unstable mode to emerge from the noise it may be already be close to
the saturation amplitude, making any estimate of the rate of growth of
the instability highly uncertain.  This behavior is illustrated in the
upper panel of Fig.~\ref{fig.amplot}, which reports the growth of a
bisymmetric disturbance in the unstable disk model described in
\S\ref{sec.modetest}.  The curves indicate very gradual progress
towards the expected result as $N$ is increased; in particular, there
was no period of exactly exponential growth even when 40 million
particles were employed.

However, \citet{Sell83} demonstrated that we can substantially reduce
the seed amplitude of all low-order non-axisymmetric instabilities by
employing the simple strategy of placing $j_\phi$ particles, each
having the same radius and velocity components in polar coordinates,
at equal intervals in $\Delta\phi = 2\pi/(m_{\rm sect}j_\phi)$ around
an arc of a circle, and starting from a randomly chosen initial phase
$0<\phi_0<\Delta\phi$ for each ring.  In this formula, if $m_{\rm
  sect}$ is the only active sectoral harmonic,\footnote{this author's
preferred term and that of \citet[][and references therein]{Lind63}
but aka angular harmonic} Fourier synthesis in azimuth replicates the
particles in other sectors. The dramatic improvement that results from
this simple change is shown in the lower panel of
Fig.~\ref{fig.amplot} that reveals approximately exponential growth by
a much larger factor that closely tracks the predicted linear growth
rate, and is not a bad match even for the smallest $N$ shown!

The minimum number of particles required per ring on the polar grid is
not quite trivial.  As ring particles are driven away from perfect
symmetry by a growing disturbance, their displacements must be
prevented from contributing, through aliases, to other force terms,
which requires the number of equally spaced particles in an arc of
$2\pi/m_{\rm sect}$ to be at least $j_\phi = (2m_{\rm max}+1)/m_{\rm
  sect}+1$.  Here $m_{\rm max}$ is the highest active sectoral
harmonic, which could be a multiple of $m_{\rm sect}$.  In more
general methods, in which force terms are not restricted to a few
low-order sectoral harmonics, the effective value of $m_{\rm sect}=1$,
and $j_\phi$ should be larger.  For a 3D axisymmetric model, $j_\phi$
should be doubled, since all particles on one ring should have the
same $z$ distance from the mid-plane and be reproduced in a reflection
symmetric ring, in which both $z$ and $v_z$ have the opposite signs.
In practice, we give each of the $j_\phi$ particles a small, typically
$\la 0.1^\circ$, random nudge in azimuth to create a seed disturbance.

This procedure suppresses shot noise in the low-order terms of the
density distribution, at the expense of a greatly enhanced signal from
the sectoral harmonic of the imposed rotational symmetry,
$m=j_\phi/(n_{\rm dim}-1)$ where $n_{\rm dim}=2$ or 3, and its
multiples.  With a polar grid, it is straightforward to suppress
forces from this artificially boosted non-axisymmetric term, but with
more general force methods, the response to the strongly enhanced
amplitude of a sectoral harmonic having $m=j_\phi$ should be mild
provided $j_\phi$ is large, say $j_\phi \ga 20$ -- see \S3.4 of
\citet{SC23} for a successful example using a Cartesian grid.

Note that the purpose of imposing initial axial symmetry is to {\em
  hide} the particulate nature of the model.  Since the gravitational
field is that of a smooth mass distribution, each of the $j_\phi$
particles pursues a congruent orbit, maintaining the initially
symmetric arrangement.  Thus the particles on each ring mimic a
circular wire of uniform mass per unit length that oscillates radially.
The rings distort smoothly as the particles respond to any developing
low-order non-axisymmetric disturbances, and generally maintain
coherence until the instability saturates.  Note that the mutual
gravitational attractions of particles are smoothed by restricting
force terms to a few active sectoral harmonics, which almost
eliminates the microscopic chaos that afflicts systems of point mass
particles \citep[\eg][]{Mill64,KS91,GHH93,HM02}.

The survival time of ring coherence depends upon the responsiveness of
the dynamical model to density fluctuations, which is a particular
problem in dynamically cool disks.  The rings break up even in the
absence of a global instability because the supporting response of the
surrounding disk to even mild density inhomogeneities results in
density wakes \citep{JT66, Binn20} that further disturb the regular
arrangement, leading to exponential growth of noise.  Fortunately, the
growth rate of an unstable mode, when present, also depends upon the
responsiveness of the disk \citep{SM22}, and generally we find that
the dominant instability outgrows the noise.  Without a dominant
instability, the growth of noise must cease when the rings have been
completely randomized, and any subsequent evolution will be no
different from that in a model started from random azimuths
\citep[\eg][]{Sell12}.

As just mentioned, the use of a polar grid to determine the gravitational
field has the further advantage that it is straightforward to restrict
non-axisymmetric force terms acting on the particles to a single
specified sectoral harmonic, $m_{\rm sect}$.  Note, however, that
simulations in which disturbance forces are restricted to a single
$m_{\rm sect}$, do not capture the correct behavior once the
instability saturates.  As the amplitude approaches saturation,
density variations develop at other sectoral harmonics, typically
$m=0$ and at low multiples of the originally active $m_{\rm sect}$,
that should contribute to the total self-consistent gravitational
field if the simulation is meaningfully to be continued beyond linear
growth.

\section{Some numerical results}
\label{sec.results}
We choose three distinct models to illustrate the advantages of the
particle selection procedures described in \S\ref{sec.smooth}.  These
models deliberately employ quite modest numbers of particles in order
to demonstrate more clearly the advantages of smooth selection.
Initial azimuthal symmetry (\S\ref{sec.qstart}) was used in only the
first example, while all three compare random sampling with smooth
selection.

\subsection{The linear mode of the isochrone disk}
\label{sec.modetest}
The improved behavior resulting from these techniques is apparent from
measurements of the frequency of a global instability.  Our chosen
example is the 2D isochrone disk \citep[][\S2.2.2(d)]{BT08}, for which
\citet{Kaln76} derived a family of DFs.  The full-mass isochrone/12
disk has $Q \simeq 1.1$ over the inner disk and \citet{Kaln78} used a
matrix method to predict the frequency of the dominant unstable mode.
We computed disturbance forces from the particles by a basis function
method \citep{ES95, Sell14}, employing $m=2$ terms only, while adding
at every step the central attraction of the disk to the
self-consistent disturbance force on each particle.  In two separate
sets of simulations, we selected 40K particles from the DF by random
sampling in one set and by smooth sampling (\S\ref{sec.2Ddisk}) in the
other set, using a different random seed for each case.  For both
sets, we imposed initial axial symmetry (\S\ref{sec.qstart}) with
$j_\phi=3$ particles spaced evenly around a half-circle (making
$N=120$K total) and ran each simulation until the instability
saturated.  The advantage of imposing axial symmetry was already
illustrated in Fig.~\ref{fig.amplot}, while here we show that smooth
selection of particles yields a further, but less dramatic
improvement.

\begin{figure}
\includegraphics[width=\hsize,angle=0]{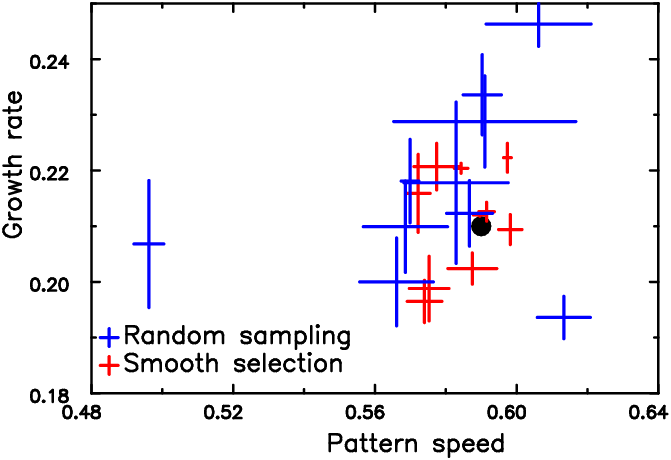}
% jasmine:misc_data/sf2d/isoc/plot.s
\caption{The scatter from many simulations of the measured pattern
  speed and growth rate of the dominant mode of the isochrone/12 disk.
  Each colored point is the best-fit value from a simulation employing
  120K particles, with error bars indicating the full range from
  different acceptable fits to the data.  The blue points record the
  measured values when random sampling is used, the red points when
  the sampling error is reduced by the smooth selection procedure
  described in \S\ref{sec.2Ddisk}.  The large black dot marks
  the frequency predicted by \citet{Kaln78} from linear theory.}
\label{fig.freqs}
\end{figure}

As usual, we measure the instantaneous amplitude of bi-symmetric
disturbances in the distribution of the $N$ particles using an
expansion in logarithmic spirals:
\begin{equation}
A(m,\gamma,t) = {1 \over N}\sum_{j=1}^N \, \exp[im(\phi_j + \tan\gamma \ln R_j)],
\label{eq.logspi}
\end{equation}
where, $m=2$, $(R_j,\phi_j)$ are the polar coordinates of the $j$th
particle at time $t$, and $\gamma$ is the (radially constant) angle of
the spiral component to the radius vector, which is the complement to
the spiral pitch angle.  We estimated the linear frequency of the
dominant mode from fits to these measurements, and to the coefficients
of the basis expansion by the method described by \citep{SA86}.

Figure~\ref{fig.freqs} presents the estimated frequency, with error
bars, of the dominant instability in each of multiple simulations
having a different random seed.  The frequency predicted from linear
theory by \citet{Kaln78} is marked by the black dot.  The blue points
report the estimated eigenfrequency in separate simulations in which
coordinates of the independent particles were selected by random
sampling, while the red points show the values from a separate set of
simulations in which the integrals for the particles were selected in
the smooth manner described in \S\ref{sec.2Ddisk}.  The blue symbols
range $\pm 21$\% (or $\pm 8\%$ if the outlier is omitted) from the
mean pattern speed and $\pm 24$\% from the mean value of the growth
rate, whereas the ranges of the red symbols are respectively $\pm
4.4$\% and $\pm 12$\%.

Thus it seems that noise in the distribution of $(E,L_z)$ values
(Fig.~\ref{fig.DFact}) causes a larger spread in the measured
frequencies than we obtained from a smooth distribution, though the
difference is not spectacular.  Differences in the measured
frequencies arise, for the most part, because the unstable mode in
each simulation is not that of the intended model, but of one having a
DF that differs by shot noise.

The error bars, and the scatter of the points, from both methods
diminish as the number of particles is increased, and indeed
\citet{ES95} were able to obtain excellent agreement with the
predicted value from larger simulations.

Figure~\ref{fig.freqs} also demonstrates that random selection the
sub-box values of $(E,L_z)$ and of radial phases described in
\S\ref{sec.2Ddisk} does not, in practice, re-establish full shot
noise, since the spread of the red points is smaller than that of the
blue.

\subsection{The rise of amplified noise in the Mestel disk}
\label{sec.Mestel}
\citet{Toom81} had predicted that the half-mass Mestel disk, which has
a constant circular speed $V_0$ at all radii, was linearly stable.
However, simulations of this model reported by \citet{Sell12} found
that swing-amplified particle noise, even with $N=5\times10^8$ particles,
created ``scratches'' in the DF as each non-axisymmetric disturbance
was absorbed through non-linear scattering at its inner Lindblad
resonance (ILR).  The modified DF then supported larger amplitude
disturbances that in turn created deeper scratches, causing secular
growth until the grooves in the DF were pronounced enough to seed a
linear instability that created a strong bar.  \citet{Sell12} selected
particles for all his simulations by the smooth method described in
\S\ref{sec.2Ddisk}, and here we show that a model in which random
sampling of particles was employed caused secular growth of
non-axisymmetric features to occur more rapidly.

We again use the 2D polar grid that \citet{Sell12} employed, with
disturbance forces restricted to $m=2$ and, as for the isochrone disk
above, we added the central attraction to each particle at each step.
We compare the evolution of two simulations having $N=5$M that
employed either smooth selection or random sampling from the DF, with
the initial azimuthal coordinates selected at random (\ie,
$j_\phi=1$).  The unit of time is $R_0/V_0$, where $R_0$ is
the mean radius of the central cutout; see \citet{Sell12} for details
of the DF, and the inner and outer tapers to the otherwise cusped and
infinite disk.

\begin{figure}
\includegraphics[width=\hsize,angle=0]{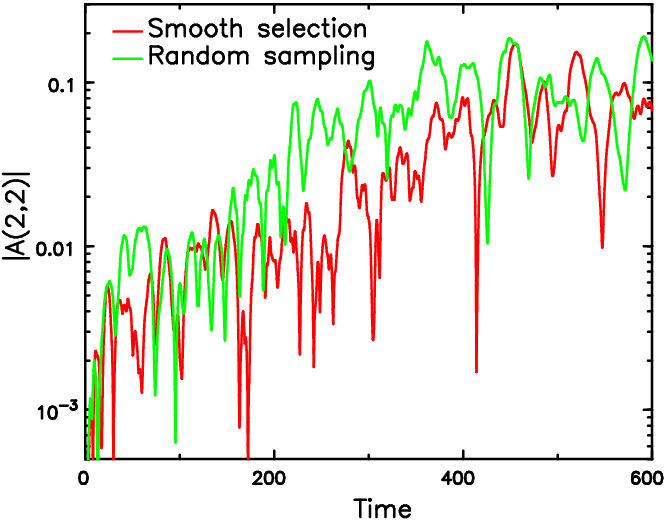}
% jasmine:/data/sellwood/5400/5459/lgsp.s 5464 5465
\caption{The time evolution of the amplitude of a logarithmic spiral
  component in two simulations of a linearly-stable, half-mass Mestel
  disk.  Both had $N=5$M particles, the same grid parameters,
  softening length, and time step.  Notice that the secular rise in
  amplitude is a little more rapid when the initial particles were
  selected by random sampling.}
\label{fig.lgsp}
\end{figure}

Figure~\ref{fig.lgsp} presents the time evolution of the $m=2$,
$\tan\gamma=2$ component in both simulations.  As \citet{Sell12}
reported, the amplitude starts out by rising quite rapidly as the
particle distribution becomes polarized, and the time evolution in
both models is characterized by rapid fluctuations due to interference
between swing-amplified transients having different corotation radii
in the disk.  Note, we stopped these simulations at $t=600$, while the
$N=5$M simulation in that earlier paper was continued to $t=1500$ with
no significant amplitude changes during the second half of the
evolution.  Note that here we report the amplitude of a logarithmic
spiral component, whereas \citet{Sell12} chose a different measure.

However, the secular rise of the amplitude in the model that began
from randomly sampled particles is more rapid than that in which
particles were selected more carefully.  The only difference between
these two cases is that values of the integrals $E$ and $L_z$ differ
by shot noise from those in the analytic DF when random sampling is
used, whereas that part of the shot noise is strongly suppressed with
smooth selection.  The more rapid secular growth is probably caused by
mild instabilities triggered by the non-uniformities in the density of
particles as a function of the integrals, of the kind illustrated in
the upper panel of Fig.~\ref{fig.DFact}.

\subsection{A disk-halo model}
\label{sec.diskhalo}
As the first model addressed an instability of an isolated disk and
the second a disk embedded in a rigid halo, we here add a slightly
more realistic example of an axisymmetric exponential disk embedded in
a halo of live particles.  The disk surface density profile is
\begin{equation}
  \Sigma(R) = {M_d \over 2\pi R_d^2} e^{-R/R_d},
\label{eq.expdisk}
\end{equation}
where $R_d$ is the disk scale length and $M_d$ is the nominal mass of
the infinite disk.  We limit its radial extent using a cubic function
to taper the surface density smoothly from $\Sigma(4.5R_d)$ to zero at
$R=5R_d$.

The halo is an originally isotropic Hernquist sphere \citep{Hern90}
which has been compresed adiabatically \citep{SM05} by the addition of
the disk mass at is center.  The original halo had a nominal mass of
$80M_d$ and a core radius of $30R_d$, and its infinite extent was
restricted by eliminating all particles having sufficient energy,
before compression, to pass $r=70R_d$, creating a smooth decrease in
density to zero at that radius.  We compress the halo using the
grid-determined, softened field of the thickened and truncated disk,
using the radial attraction in the disk mid-plane, which we adopt as
spherically symmetric.  Since both the angular momentum and radial
action are conserved during compression, the compressed $f_c(E,L)$,
which becomes anisotropic, has the same value for the two actions
$J_r$ and $J_\phi$ \citep{Youn80, SM05} as the original $f_H(E)$ given
in Hernquist's paper.

\begin{table}
\caption{Numerical parameters of the disk-halo simulations} % run5438
\label{tab.pars}
\begin{tabular}{@{}ll}
Cylindrical polar grid \\
\qquad Mesh points in $(r, \phi, z)$ & 127 $\times$ 192 $\times$ 125  \\
\qquad Spacing of planes & $R_d/50$ \\ 
\qquad Active sectoral harmonics & $m = 0$ only \\
\qquad Spline softening length & $\epsilon = R_d/20$ \\
Spherical grid \\
\qquad Outer boundary & $r_{\rm max} = 80R_d$ \\
\qquad Radial shells & 501 \\
\qquad Active surface harmonics & $l = 0$ only \\
Grid scaling & $R_d = 10$ grid units \\
Number of disk particles & $2 \times 10^5$ \\
Number of halo particles & $2 \times 10^5$ \\
Shortest and longest time-step & $\tau_{\rm dyn}/80$ and $\tau_{\rm dyn}/5$ \\
Radial time step zones & 5 \\
\end{tabular}
\end{table}

Once the combined potential, $\Phi(R,z)$, of the disk and compressed
halo is known, we can solve for the DF of the disk, using the method
proposed by \citet{Shu69}.  The in-plane DF has the form
\begin{equation}
  f(E,L_z) = \cases{{\cal F}(L_z) e^{-{\cal E}/\sigma_R^2(R_g)} & $0<{\cal
    E} \leq -E_c(L_z)$, \cr 0 & $L_z <0$. \cr}
\label{eq.ShuDF}
\end{equation}
Here ${\cal E}$ is the excess energy of a particle above $E_c$, which
is that of a circular orbit at the guiding center radius $R_g(L_z)$.
Although this DF assumes no retrograde stars, we later reverse the
angular momentum of some low-$L_z$ particles in order to smooth the
discontinuity in $f(E,L_z)$ at $L_z=0$, which does not affect the
equilibrium.  The DF (\ref{eq.ShuDF}) clearly assumes a Gaussian
radial velocity distribution everywhere, but the azimuthal velocity
distribution is appropriately skewed.  We set the radial velocity
dispersion of the disk particles using the \citet{Toom64} criterion
$\sigma_R(R) = Q \; 3.36 G\Sigma/\kappa$, where $\kappa$ is the local
epicyclic frequency \citep{BT08}, choosing $Q=1.5$ at all radii.

The function ${\cal F}(L_z)$ has to be determined numerically and the
procedure we adopt is described in the on-line manual \citep{Sell14}.
As there are many possible functions ${\cal F}$ that fit the adopted
disk surface density and $Q$ profile, we impose two extra
requirements.  Not only are rapid fluctuations of ${\cal F}$ with
$L_z$ physically unreasonable, but we have also found that even mild
``ripples'' in the function ${\cal F}(L_z)$ can seed disk
instabilities related to groove modes \citep{SM22}.  We therefore
penalize the fit to the surface density also to minimize $T =
\sum_{L_z}[ {d^2{\cal F} / dL_z^2}]^2$.  Note that the numerical
search for the optimum ${\cal F}$ seeks a balance between fitting the
disk surface density while also minimizing $T$, and finding the
optimum balance is something of an art.  It is also required that
${\cal F}(L_z) \geq 0$ for all $L_z$, although we find that this
requirement is generally satisfied for a smooth ${\cal F}$ without
imposing an additional constraint.

The disk particles are distributed vertically in a Gaussian
distribution having a spread $z_0 = 0.1R_d$ at all radii and their
vertical velocities are set using eq.~(\ref{eq.vert1D}).

Here we use units such that $G=M_d=R_d=1$.  Our unit of time
is therefore $\tau_{\rm dyn}=(R_d^3/GM_d)^{1/2}$.  For those who
prefer physical units, a possible scaling is to set $R_d=2\;$kpc, and
$\tau_{\rm dyn}=10\;$Myr, which implies $V_0 \simeq 176\;$km~s$^{-1}$
and $M_d \simeq 1.78\times 10^{10}\;$\Msun.

\begin{figure}
\includegraphics[width=\hsize,angle=0]{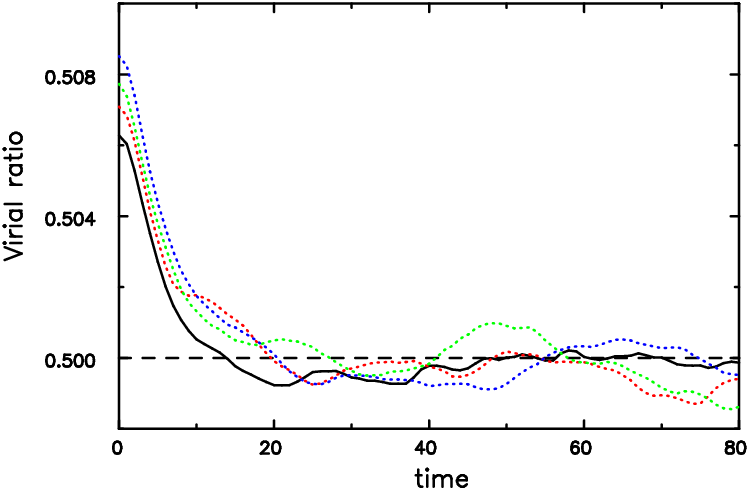}
% jasmine:/data/sellwood/5400/5442/analys.s
\caption{The evolution of the virial ratio of multiple realizations of
  the model described in \S\ref{sec.diskhalo}.  The solid black line
  is for a simulation in which both the disk and halo particles were
  selected in the smooth manner described respectively in \S\S\ref{sec.2Ddisk} \&
  \ref{sec.sphrd}.  The colored dotted curves are from simulations in
  which the initial coordinates of both components were chosen by
  random sampling, using a different random seed in each case.  The
  very slight initial disequilibrium, reflected by the initial drop,
  is caused by the vertical adjustment of the disk thickness, because
  of the approximate set up using eq.~(\ref{eq.vert1D}).  All four
  simulations were forced to remain axisymmetric for this test to
  prevent the development of a bar, which would otherwise begin to
  emerge by $t\sim30$.}
\label{fig.virplt}
\end{figure}

We compute the evolution of the model using a hybrid grid method, see
\citet{Sell14} for a full description.  The evolution of the disk
particles is computed using a 3D cylindrical polar grid, while
accelerations of the halo particles are computed through a spherical
grid; naturally, all particles experience forces from those of both
components at every step.  The numerical parameters adopted are
given in Table~\ref{tab.pars}.

Figure~\ref{fig.virplt} reports the evolution of the virial ratio,
$T/|W_C|$, over a short time interval of $80\tau_{\rm dyn}$, or about
three orbit periods at $R=3R_d$, in a number of simulations.  The
virial of Clausius $W_C = \sum_N \mu_i \ba_i \cdot \bx_i$, with
$\mu_i$, $\ba_i$, and $\bx_i$ being respectively the mass,
acceleration, and position of the $i$th particle at the given moment.
The model is very slightly out of balance initially, as $T/|W_C| \sim
0.506$ (note the small range of the vertical scale), but settles
within $\sim 20\tau_{\rm dyn}$ as the disk adjusts to its very mild
disequilibium caused by the imperfection of eq.~(\ref{eq.vert1D}).

The solid black line reports the case in which particles were selected
from the DFs for both the disk and halo using the smooth methods
described in \S\ref{sec.smooth}, and the colored dotted lines are from
models that used random sampling, each having a different random seed.
Since the virial ratio is affected in later evolution by a growing bar
instability in this disk-halo model, we suppress all non-axisymmetric
force terms in order that the model simply settles to an axisymmetric
equilibrium.  It can be seen from this Figure that random sampling not
only introduces differences in the initial virial ratio, but causes
persisting fluctuations that reflect mild collective changes from
noise-driven evolution.  However, the differences are minor.

\section{Conclusions}
\label{sec.concl}
In order to construct a simulation that is initially in equilibrium,
it is best to select particle coordinates from a distribution function
(DF).  Having invested effort to create an equilibrium DF, it makes
sense also to try to minimize the sampling error when selecting
particles from it.  We have presented methods that can be used to
reduce sampling noise in the selection of particles and, though they
require extra effort to program, they do not take significantly more
cpu time to run than does the random sampling approach that is widely
used.  With random sampling, initial noise can also be reduced, albeit
slowly, by increasing the number of particles, but methods that yield
a similar improvement without increasing $N$ have clear practical
advantages.

We have presented three sets of models to illustrate the advantages of
this approach.  These simulations employ quite modest numbers of
particles in order to illustrate the advantages more clearly.  The
frequency of an unstable mode is more reliably reproduced when
particles are selected in a smooth manner in the case examined in
\S\ref{sec.modetest}, while we show in \S\ref{sec.Mestel} that
non-axisymmetric disturbances grow slightly more rapidly in
simulations of the stable Mestel disk when started from a randomly
sampled set of particles.  Finally, the models presented in
\S\ref{sec.diskhalo} are slightly closer to equilibrium when smooth
sampling is employed.  The improvement in all three cases is clear,
though not dramatic.

\section*{Acknowledgements}
The ideas described here were developed over many years in
collaborations with Lia Athanassoula, Victor Debattista, David Earn,
Juntai Shen, and Monica Valluri, and I am eternally grateful for their
assistance.  I am especially indebted to Victor Debattista, who
suggested I include the example of the Mestel disk presented in
\S\ref{sec.Mestel}.  An anonymous referee made a number of thoughtful
suggestions to strengthen the paper.  The author also acknowledges a
helpful email discussion with James Binney and the continuing
hospitality and support of Steward Observatory.

\section*{Data availability}
The data from the simulations reported here can be made available on
request.  The source code for particle selection, as well as the
simulation and analysis software can be downloaded in one bundle from
{\tt http://www.physics.rutgers.edu/galaxy}, and is fully
documented in the code manual \citep{Sell14}.

\bsp	% typesetting comment
\label{lastpage}
\end{document}